\documentstyle[11pt,epsfig]{article}
\begin{document}

\begin{center}
{\bf On a SUSY QM scheme for any linear homogeneous differential equation of the
second order}
\end{center}

\bigskip

\begin{center}
R. KLIPPERT$^{1,2}$ and H.C. ROSU$^{1,3}$ 


{\scriptsize $^1$ International Center for Relativistic Astrophysics, 
Piazzale della Repubblica 10,\\[-1.2ex] 65100 Pescara, Italy}

{\scriptsize $^2$ Brazilian Center for Research in Physics, R. Dr.
Xavier Sigaud 150 Urca, 22290-180 \\[-1.2ex]Rio de Janeiro RJ, Brazil}\\[-0.7ex]

{\scriptsize $^3$ Instituto de F\'{\i}sica, Univ.\ de Guanajuato, 
Apdo Postal E-143, Le\'on, Gto, Mexico}

\end{center}

\bigskip
\bigskip

\noindent
{\bf Summary.} - A formal supersymmetric quantum mechanics (SUSY QM)
 procedure for any linear homogeneous
second-order differential equation is briefly sketched up and applied 
to a simple exactly solvable case. 

\bigskip 
\bigskip 

\bigskip 
\bigskip 
\bigskip

It is quite well known how to get Riccati equations from the general form of
homogeneous linear differential equation of the second order \cite{1}

\begin{equation} \label{e1}
A(x)\frac{d^2 u}{dx^2} +B(x)\frac{du}{dx}+ C(x)u=0~.
\end{equation}

It requires the transformation 

\begin{equation} \label{e2}
y=\frac{1}{R(x)}\frac{1}{u}\frac{du}{dx}~, 
\end{equation}
where $R(x)$ is an arbitrary function. Using $u^{'}=Ruy$ and 
$u^{''}=Ru(y^{'}+Ry^{2}+\frac{R^{'}}{R}y)$ in Eq.~(\ref{e1}) leads to

\begin{equation} \label{e3}
\frac{dy}{dx}+\left(\frac{R^{'}}{R}+\frac{B}{A}\right)y+Ry^2=-\frac{C}{AR}~.
\end{equation}

Moreover, since $R(x)$ is an arbitrary function, one can choose it such that 
the coefficient of $y$ be zero, i.e., one can end up with the simplified
Riccati equation

\begin{equation} \label{e4}
\frac{dy}{dx}+ R_0y^2=-\frac{C}{AR_0}~, 
\end{equation}
where $R_{0}(x)=\exp[-\int ^{x}(B/A)dx^{'}]$.

Our point here is to notice that Eq.~(\ref{e4}) allows a direct 
connection with various SUSY QM 
schemes \cite{HR} by means of the change of independent variable 
$z(x)=\int ^{x}R_{0}dx^{'}$. This 
leads to a Riccati equation of SUSY QM type at zero factorization energy
along the curve $z(x)$
\begin{equation} \label{e5}
\frac{dy}{dz}+y^2=V_{1}(z)~,
\end{equation}
where $V_{1}(z)=-\frac{C(z)}{A(z)R_0^2(z)}$ can be interpreted as a 
Schr\"odinger `potential'. The SUSY partner Riccati equation will be 
\begin{equation} \label{e6}
-\frac{dy}{dz}+y^2=V_{2}(z)~,
\end{equation}
where the partner `potential' $V_{2}(z)$ is Darboux `isospectral' with 
respect to $V_{1}$, i.e.
\begin{equation} \label{e7}
V_{2}=V_{1}-2D^2[\ln (\psi(z))]~,
\end{equation} 
where $D=\frac{d}{dz}$ and $\psi$ is a particular solution of $D^2\psi-V_1\psi=0$.

In addition, one can think of the SUSY QM scheme based on the general 
Riccati
solution, as first tackled by Mielnik for the harmonic oscillator 
case \cite{M}. In the latter approach,
one gets a one-parameter family of Darboux strictly isospectral
`potentials' given by
\begin{equation} \label{e8}
V_{1}(z;\lambda)=V_{1}-2D^2[\ln (I(z)+\lambda)]~,
\end{equation}
where $I(z)=\int ^{z}\psi^2(t)dt$. Moreover,
\begin{equation} \label{e9}
\psi(z,\lambda)=\frac{\psi(z)}{\int ^{z}\psi^2(t)dt+\lambda}
\end{equation}
is the modulated Schr\"odinger zero mode implied by this scheme in which 
the Riccati integration constant $\lambda$ is kept as a free parameter. 
In general,
to get continuous solutions $\psi(z,\lambda)$, one should take care of not
having a zero denominator. This leads to conditions on the possible values 
of $\lambda$.
On the other hand, if one works with polynomial solutions there will be 
singularities in the log derivative of $\psi$. SUSY partner potentials based 
on the $n^\prime$th excited state of a Schr\"odinger discrete spectrum problem
split in $n+1$ branches separated by the $n$ singularities of the log 
derivative \cite{Rob}.

Even more general schemes, such as higher-order intertwinings can be applied 
leading to a rich class of `isospectral' solutions.

The main drawback of the formal scheme as presented in 
Eqs.~(\ref{e1})--(\ref{e7})
is that it is not at all easy to implement in practice because onto the $z$ axis one 
does not get an easily solvable eigenvalue problem.
Let us take as an example the Hermite equation
\begin{equation} \label{e10}
u^{''}-2xu^{'}+2nu=0~.
\end{equation}
For this case, $A(x)=1$, $B(x)=-2x$, and $C(x)=2n$. Thus, 
$R_{0}(x)={\rm e}^{x^2}$
and $z(x)=\int ^{x}{\rm e}^{t^2}dt$. The potential $V_1(z)=-2n/{\rm e}^{2z^2}$ does not lead
to an exactly solvable problem. 

We provide in the following an exactly solvable case focusing on the scheme
based on the general Riccati solution. Let us take the initial equation of 
the form
\begin{equation} \label{e11}
D_{x}^{2}u+ (1/x)D_xu+u=0
\end{equation} 
Then $V_1=-z^2$ that corresponds to an 
eigenvalue problem in terms of the Bessel functions, namely $\psi(z)=C_1\sqrt{z}J_{\frac{1}{4}}
(z^2/2)+C_2\sqrt{z}Y_{\frac{1}{4}}(z^2/2)$.
In Fig.\ 1 we present a common three-dimensional plot of the formulas (\ref{e8}) 
and (\ref{e9}) for the pair of superposition constants $C_1=0$ and $C_2=1$.
On the other hand, in Fig.\ 2 we present a full axis one-dimensional plot 
for $\lambda =0.2$ and $C_1=1$ and $C_2=0$.

One can also ask what is the equation in the initial $x$-axis corresponding
to the potential $V_{1}(z,\lambda)$. 
One can easily show that such an equation has the following form
\begin{equation} \label{e12}
D_{x}^{2}\psi\Big(z(x),\lambda\Big)+ \frac{1}{x}D_x\psi\Big(z(x),\lambda\Big)-\frac{V_{1}(z,\lambda)}{x^2}\psi\Big(z(x),\lambda\Big)=0~.
\end{equation}
For this case, the connection between the two axes is given by
$z={\rm ln}x$.

\begin{figure}[htbp]
\leavevmode
\centerline{
\centering
\epsfxsize=80ex
\epsfbox{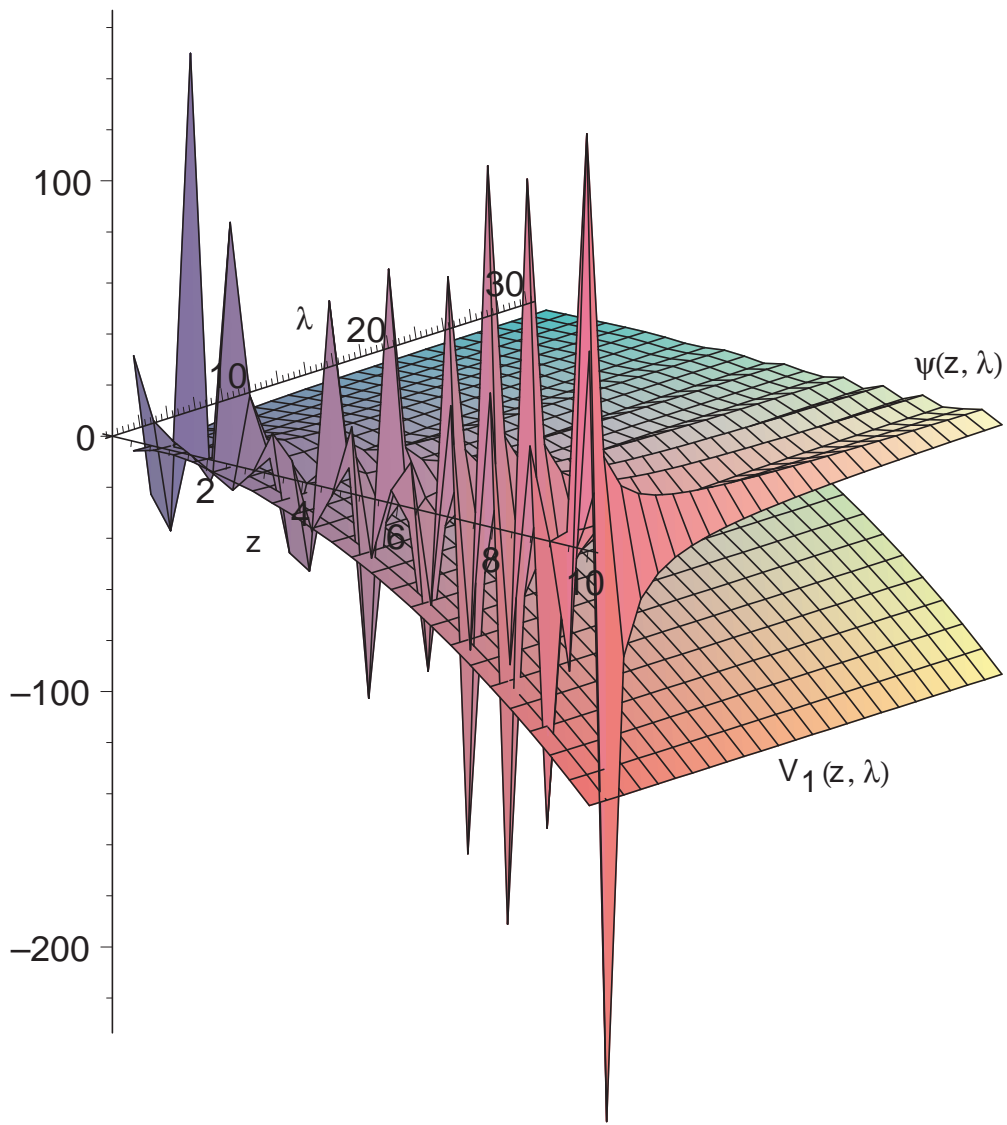}}
\caption{The potential $V_1(z,\lambda)$ and the eigenfunctions 
$\psi(z,\lambda)$ corresponding to $\psi(z)\sim\sqrt{z}\,Y_{\frac{1}{4}}(z^2/2)$ for $\lambda \in (0,30)$.  
The dependence on $\lambda$ is almost flat in the range of large $\lambda$.}
\label{fig1}
\end{figure}

\begin{figure}[htbp]
\leavevmode
\centerline{
\centering
\epsfxsize=60ex
\epsfbox{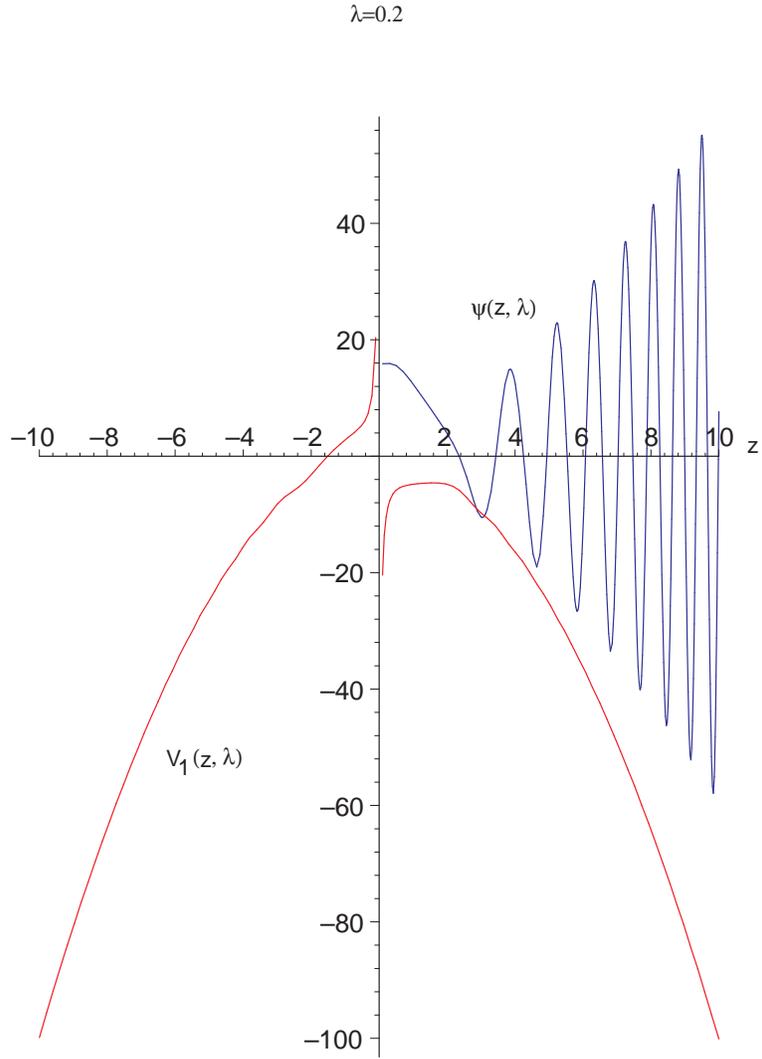}}
\caption{The eigenfunction $\psi(z,\lambda)$ corresponding to $\psi(z)\sim\sqrt{z}J_{\frac{1}{4}}(z^2/2)$ 
and the potential $V_1(z,\lambda)$ for $\lambda=0.2$.  
The wave function seems to approach a finite value for $z=0$, 
which can be explained by the presence of the opposite divergences 
in the potential $V_1(z,\lambda)$ at $z=0$.}
\label{fig2}
\end{figure}

\clearpage


\section*{Acknowledgements}

\noindent
We thank Prof. R. Ruffini for hospitality at ICRA-Pescara.
RK would like to thank Brazilian CAPES founding agency for a grant.

\end{document}